\documentclass{elsarticle}

\usepackage{mathtools}

\begin{document}

\begin{frontmatter}

\title{Queueing systems with renovation vs. \break queues with RED. Supplementary Material.}

\author{Mikhail Konovalov}
\address{Institute of Informatics Problems of the FRC CSC RAS, Moscow, Russia}
\ead{mkonovalov@ipiran.ru}

\author{Rostislav Razumchik}
\address{Institute of Informatics Problems of the FRC CSC RAS, Moscow, Russia \\  Peoples' Friendship University of Russia (RUDN University), Moscow, Russia}
\ead{rrazumchik@ipiran.ru, razumchik\_rv@rudn.university}

\begin{abstract}
In this note we consider $M/D/1/N$ queue with renovation
and derive analytic expressions for the following
performance characteristics:
stationary loss rate, moments of the number in the system.
Moments of consecutive losses, waiting/sojourn time (as introduced
in \cite{Bonald}) are out of scope.  
The motivation for studying these characteristics is in 
the comparison of renovation with 
known active queue mechanisms like RED.
\end{abstract}


\end{frontmatter}


\section{System description}

We consider the system consisting of one queue of finite capacity $N$,
served by single server. 
Customers arrive at the system according to the Poisson flow of rate $\lambda$.
If a customer sees the system full it is lost,
otherwise it occupies one place in the queue if the server
is busy and the server if it is idle.
Service times are constant, equal to $d>0$.
Thus the cumulative distribution function $B(x)$ 
of the random variable equal to customers service time is
the step function: it is $0$ if $x<d$ and $1$ otherwise.
Upon service completion one customer from the head
of the queue enters server i.e. the service discipline is FCFS.

General renovation mechanism is implemented in the system.
It works as follows. Define $N+1$ numbers, say $q_i\ge 0$,
$0 \le i \le N$, satisfying \mbox{$\sum_{i=0}^N q_i=1$}.
When the service is completed the served customer leaves the system
and additional $i$ customers are removed from the 
queue with probability $q_i$. Such mechanism of removing
customers from the system is called renovation 
in \cite{Zaryadov2010,Kreinin}.

For the purpose of comparison with RED
we need to introduce several refinements into the renovation procedure. 
Firstly notice that after the renovation, the queue may become empty 
and thus the server will be idle until the next arrival. From the practical point of view 
it is more appealing to leave at least one customer in 
the system after the renovation.
Secondly, it may happen that upon service completion it is required to remove
more customers, than are actually waiting in the queue.
For such a conflict we will consider separately two resolution options:

\noindent 
\textit{Option 1}. 
If upon service completion there are $1 \le i \le N$ customers
waiting in the queue, then 
\begin{itemize}
\item[--] with probability $q_0$ nothing happens;
\item[--] with probability $q_j$, $0<j<i$, exactly $j$ customers
from the queue leave the system and those customers 
are chosen successively \textit{starting from the head of the queue};
\item[--] with probability $Q_i=q_i+q_{i+1}+\dots+ q_N$ exactly $(i-1)$
customers from the queue leave the system. Again those customers 
are chosen successively \textit{starting from the head of the queue}.
\end{itemize}

\noindent \textit{Option 2}. 
If upon service completion there are $1 \le i \le N$ customers
waiting in the queue, then 
\begin{itemize}
\item[--] with probability $q_0+Q_i$ nothing happens, where 
$Q_i=q_i+q_{i+1}+\dots+ q_N$;
\item[--] with probability $q_j$, $0<j<i$, exactly $j$ customers
from the queue leave the system and those customers 
are chosen successively \textit{starting from the head of the queue}.
\end{itemize}

\noindent Throughout the paper, for the sake of brevity,
we use the agreement that $\sum_{k=0}^{-1} \equiv 0$.

\section{Option 1}

\subsection{Stationary distribution}

Let $N(t)$ be the total number of customers at instant $t$
and $E(t)$ be the elapsed service time of the customer in server
(if there is one) in the $M/D/1/N$ queue with renovation as described above.
In order to compute the stationary queue size moments we need the distribution
\begin{equation}
\label{pn}
\lim_{t \rightarrow \infty} \mathbf{P}\{ N(t)=n \}=P_n,\  0 \le n \le N+1,
\end{equation} 
and for the loss rate, due to PASTA property of Poisson arrivals, -- the distribution
it is sufficient to know
\begin{equation}
\label{pnx}
\lim_{t \rightarrow \infty} \mathbf{P}\{ N(t)=n, E(t)<x \}=P_n(x), \ 1\le n \le N, \ x \in [0,d],
\end{equation}
Since we are dealing with the finite-capacity 
queue and work conserving service discipline, these 
stationary distributions exist.  
The analytic method for finding $P_n$ 
has been developed in \cite{Zaryadov2009,Zaryadov2010}.
Even though the renovation mechanism that we consider 
differs from the one in \cite{Zaryadov2009,Zaryadov2010},
the method still works. 
The distributions (\ref{pn}) and (\ref{pnx}) can be found as follows. 
At first we find the stationary distribution \mbox{$\{P^+_n, \ 0 \le n \le N\}$}
of the Markov chain $\{ \nu(t), \ t \ge 0\}$ embedded at service completion 
epochs. 
Then using well-known results for the Markov regenerative processes (see \cite[Theorem 9.19]{kulk}),
we calculate \mbox{$\{P_n, \ 0 \le n \le N+1\}$} by $P_n=\sum_{i=0}^N P^+_i f_{in} / f^*$,
where $f_{in}$ is the mean time spent by the system in state $n$, starting
from $i$, and $f^*$ is the mean time between transitions of the $\{ \nu(t), \ t \ge 0\}$.
Finally, relations for the functions $P_n(x)$ are found from the results for
the $M/D/1/N$ queue.

Let $\beta_i=[{(\lambda d)^i / i!}]e^{-\lambda d}$ and
$B_0=1-\beta_0$, $B_i=B_{i-1}-\beta_i$, $i\ge 1$. 
The entries of the transition probability matrix $P=(p_{ij})$
of the embedded Markov chain $\{ \nu(t), \ t \ge 0\}$ 
have the form
$$
p_{0j}=p_{1j}=
\begin{cases}
\beta_0, & j=0,\\
\sum_{k=1}^N \beta_k q_{k-1} +  B_N q_{N-1}+
\\
\,\,\,\,\,\,\,\,\,\,\,\,\,\,\,\,\,\,+ \sum_{k=1}^N \beta_k Q_{k} +  B_N q_{N}, & j=1,\\
\sum_{k=j}^N \beta_k q_{k-j} +  B_N q_{N-j}, &  2 \le j \le N,
\end{cases}
$$
$$
p_{ij}=
\begin{cases}
0, & j=0,\\
\sum_{k=0}^{N+1-i} \beta_k Q_{k+i-1} +  B_{N+1-i} q_{N}
+
\\
\,\,\,\,\,\,\,\,\,\,\,+ 
\sum_{k=\max(0,2-i)}^{N+1-i} \beta_k q_{k+i-2} +  B_{N+1-i} q_{N-1}, & j=1,\\
\sum_{k=\max(0,j-i+1)}^{N+1-i} \beta_k q_{k-j+i-1} +  B_{N+1-i} q_{N-j}, & 2 \le j \le N,\\
\end{cases}
\ \ 2 \le i \le N.
$$
The matrix $P=(p_{ij})$ does not have any special structure
and so the values of $P_n^+$ are found by solving the system
of linear algebraic equations 
$$
{\vec P}^+={\vec P}^+P, \ \
{\vec P}^+ {\vec 1} =1,
$$
where ${\vec P}^+= (P^+_0,\dots,P^+_N)$. 
There are numerous methods for performing the solution
(for example, Gaussian elimination method. For others one can 
refer to \cite{stew}).

In order to compute the values of
\mbox{$\{P_n, \ 0 \le n \le N+1\}$} using the relation
 $P_n=\sum_{i=0}^N P^+_i f_{in} / f^*$,
we need expressions for $f_{in}$ and $f^*$.
For the stationary mean time $f^*$
between transitions of the embedded Markov chain $\{ \nu(t), \ t \ge 0\}$
it holds that
$$
f^*=P^+_0 \left ( {1\over \lambda} + d \right )
+ (1-P^+_0) d.
$$

\noindent Since customers, that are waiting in the queue,
can leave the queue \textit{only} on service completions,
we have
$$
f_{in}
=
\begin{cases}
0, & 0 \le n \le i-1,\\
f_{0,n-i+1}, & i \le j \le N+1,
\end{cases}
\ \ 1 \le i \le N.
$$
Clearly $f_{00}={1/ \lambda}$
and other values of $f_{0n}$ are computed by conditioning on the number of arrivals
during one service completion. 
For $f_{01}$ we can write
$$
f_{01}=\int_0^\infty x e^{-\lambda x} dB(x)
+ 
\int_0^\infty dB(x) \int_0^x t \lambda e^{-\lambda t} dt, 
$$
which, by remembering the property of the Laplace-Stieltjes 
transform \mbox{$\int_0^\infty g(x) dB(x)=g(d)$},
can be reduced to 
$$
f_{01}= d e^{-\lambda d} 
+ 
{1\over \lambda} \left ( 1- \sum_{k=0}^1 {(\lambda d)^k \over k!} e^{-\lambda d} \right ). 
$$
It is straightforward to generalize this result for $1 \le n \le N$:
$$
f_{0n}=  {1\over \lambda} {(\lambda d)^{n} \over n!} e^{-\lambda d}
+ 
{1\over \lambda} \left ( 1- \sum_{k=0}^n {(\lambda d)^k \over k!} e^{-\lambda d} \right )
=
{1\over \lambda} \left ( 1- \sum_{k=0}^{n-1} {(\lambda d)^k \over k!} e^{-\lambda d} \right ). 
$$
For $f_{0,N+1}$ the expression will be different, which is due to the fact that 
the system capacity is finite and at some instant (when the queue becomes full)
the state of the system stops changing due to new arrivals 
and will change only when the service is completed. 
One way to compute  $f_{0,N+1}$  is to consider all possible events, which gives
$$
f_{0,N+1}=
\int_0^\infty x \left ( 1- \sum_{k=0}^{N-1} {(\lambda x)^k \over k!} e^{-\lambda x} \right ) dB(x)
+ 
\int_0^\infty 
{N \over \lambda} 
\left ( \sum_{k=0}^{N} {(\lambda x)^k \over k!} e^{-\lambda x} -1 \right )
dB(x) 
=
$$
$$
=
d \left ( 1- \sum_{k=0}^{N-1} {(\lambda d)^k \over k!} e^{-\lambda d} \right ) 
+ 
{N \over \lambda} 
\left ( \sum_{k=0}^{N} {(\lambda d)^k \over k!} e^{-\lambda d} -1 \right ).
$$

\noindent The other way is to recall that $\sum_{n=1}^{N+1}f_{in}=d$
and thus once $f_{0n}$, $1 \le n \le N$, are computed, $f_{0,N+1}=d-\sum_{n=1}^{N} f_{in}$.
Since $P_n$ are found, the moments $\mathbf{E} N^m$ of the total number in the system
can be computed according to the definition i.e.
$\mathbf{E} N^m
=
\sum_{k=0}^{N+1} k^m p_k$.

Coming back to the functions $P_n(x)$, notice that
the differential equations for the functions $p_n(x)=P'_n(x)$ coincide with those for 
the classical $M/D/1/N$ queue. Thus 
they have the form (see, for example, \cite[subsection 4.14]{Riordan1962}):
\begin{equation}
\label{eq3}
p_n(x)=e^{-\lambda x} [1-B(x)] \sum\limits_{k=0}^{n-1} p_{n-k}(0) {(\lambda x)^k \over k!}, \ 1 \le n \le N.
\end{equation}
Here $p_{n}(0)$ are the boundary conditions, which are in our case different 
from the boundary conditions for the classical $M/D/1/N$ due to the presence of renovation.
We can follow the classic argumentation for obtaining the boundary conditions for
$M/G/1$-type queues remembering renovation.
But since the probabilities $P_n=\int_0^d p_n(x) dx$ have been found above, 
we can integrate (\ref{eq3}) from $0$ to $\infty$ and find the 
relation between $P_n$ and $p_{n}(0)$.  This gives
\begin{eqnarray}
P_n &=& \sum\limits_{k=0}^{n-1} {\lambda^k \over k!} p_{n-k}(0)
\int_0^d e^{-\lambda x} [1-B(x)] x^k  dx
=
\nonumber
\\
&=&
{1\over \lambda}
\sum\limits_{k=0}^{n-1} B_k p_{n-k}(0), \ 1 \le n \le N. 
\label{s1}
\end{eqnarray}
This is the system of $N$ linear algebraic equations
with $N$ unknowns $p_1(0), \dots, p_N(0)$, which can be solved 
iteratively, starting from $n=1$:
\begin{eqnarray}
p_n(0)
=
{1 \over B_0}
\left (
\lambda P_n
- 
\sum\limits_{k=1}^{n-1} B_k p_{n-k}(0)
\right ), \ 1 \le n \le N.
\end{eqnarray}
Since the values of $p_n(0)$, $1 \le n \le N$,
are now known and the functions $p_n(x)$ can be computed from (\ref{eq3}).

\subsection{Loss probability}

Let $\pi$ be the probability that the arriving customer (or tagged customer)
will be lost. Due to the PASTA property of Poisson arrivals,
$p_n(x)$ is also the probability density that the arriving 
customer sees $n$ customers in the system and sees the elapsed
service time equal to $x$. 
Firstly, the arriving customer is lost
if it sees the system full, which happens with probability
$P_{N+1}$.
If the arriving customer sees the system busy but not full, 
then the derivations become tricky.
Indeed, let the arriving customer see one customer in the system 
and the elapsed service time equal to $x$ (the probability density of
this even is $p_1(x)$).
Then if no customers arrive until the service is completed 
(i.e. during the time $d-x$), then the tagged customer will
not be lost. But if there was at least one arrival
during the remaining service time $d-x$, then
the tagged customer will be lost with probability $Q_1$. 
Now assume the tagged customer sees  one customer 
waiting in the queue and the elapsed service time equal to $x$ 
(the probability density of this even is $p_2(x)$).
Then if no customers arrive until the current service is completed 
(i.e. during the time $d-x$), then the tagged customer
will be lost with some (yet unknown) probability $r_0$.
But if there was at least one arrival during the remaining service time $d-x$, then
the tagged customer will be lost with some (also unknown) probability $r^*_0$.
Clearly $r_0$ ($r^*_0$) is the probability of the event ``the customer is lost if
there are 0 customers in front of it in the queue and no  customers (at least one customer) behind it
and the elapsed service time of the customer in server is 0''
and thus
$$
r_{0} = [1-e^{-\lambda d}] Q_1,
\ \
r^*_{0} = Q_1.
$$
For the tagged customer seeing $i$ customers 
waiting in the queue and the elapsed service time equal to $x$,
we will have two probabilities $r_{i-1}$ and $r^*_{i-1}$,
which are computed recursively from $r_{j}$ and $r^*_{j}$, $0\le j<i-1$.
Putting it altogether, 
we have the following expression for the loss probability $\pi$:
\begin{eqnarray}
\pi= P_{N+1}+ Q_1 \int_0^d p_1(x) [1-e^{-\lambda (d-x)}] dx  
+
\sum\limits_{i=2}^{N-1}
\int_0^d p_i(x) e^{-\lambda (d-x)} dx \sum\limits_{j=0}^{i-2} q_j r_{i-2-j}
+
\nonumber
\\
+
\sum\limits_{i=2}^{N-1}
\int_0^d p_i(x) [1-e^{-\lambda (d-x)}] dx 
\left (
\sum\limits_{j=0}^{i-2} q_j r^*_{i-2-j}+Q_i
\right )
+
P_N \sum\limits_{j=0}^{N-2} q_j r_{N-2-j},
\label{ploss}
\end{eqnarray}
where the probabilities $r_{i}$ and $r^*_{i}$ are computed
from relations
\begin{eqnarray}
r_{0} \!\!\!\!\!&=&\!\!\!\!\! [1-e^{-\lambda d}] Q_1,
\\
r^*_{0} \!\!\!\!\!&=&\!\!\!\!\! Q_1,
\\
r_{i} \!\!\!\!\!&=&\!\!\!\!\!
e^{-\lambda d} \sum\limits_{j=0}^{i-1} q_j r_{i-1-j}
\!+\!
[1\!-\!e^{-\lambda d}] 
\left (
\sum\limits_{j=0}^{i-1} q_j r^*_{i-1-j}\!+\!Q_{i+1}
\right )\!\!,  1 \le i \le N\!-\!2,
\\
r^*_{i} \!\!\!\!\!&=&\!\!\!\!\!
\sum\limits_{j=0}^{i-1} q_j r^*_{i-1-j}\!+\!Q_{i+1},  1 \le i \le N\!-\!2. 
\end{eqnarray}
Even though the expression (\ref{ploss}) can be simplified by computing 
the integrals explicitly, it is not our goal here.
For small and moderate values of $d$, $N$ and $\lambda$
the expression (\ref{ploss}) presents almost no computational
difficulties and can be directly used for numerical implementation.

\subsection{Consecutive losses}

We will not derive here the expression for the moments of the consecutive losses
and just notice the following. When comparing with RED-type
schemes, we are interested in consecutive losses between two accepted arrivals.
Due to the fact that losses of two types occur in the system (due to the full queue and due to the renovation)
this is a hard nut to crack.
One of the feasible solutions follows from the results for the distribution 
of consecutive losses under a RED scheme. In order to make the exposition 
simpler and the argument more transparent, 
we change the deterministic service to exponential with rate $\mu$ (thus we deal
in this subsection with the $M/M/1/K$ queue).
Let $L$ be the random variable equal to a length of a series 
of consecutive losses.
The probability $\mathbf{P}\{ L=k \}$ is equal to the fraction
\begin{equation}
\label{cl1}
{\mathbf{P} \{\mbox{``an arrival accepted, next $k$ arrivals lost, the next arrival accepted''} \}
\over
\mathbf{P} \{\mbox{``an arrival accepted, the next arrival lost''} \}
}.
\end{equation}
Given that the system is busy, the probability that an arrival occurs earlier that the 
service completion is simply $\delta=\lambda/(\lambda+\mu)$.
Again by utilizing the PASTA property of Poisson arrivals
and the law of total probability we obtain the expression for 
the denominator in (\ref{cl1}):
\begin{multline*}
\mathbf{P} \{\mbox{``an arrival accepted, the next arrival lost''} \}
=
\\
=
\sum_{n=0}^N P_n (1-d_{n})\sum_{i=1}^{n+1} \delta d_i (1-\delta)^{n+1-i}.
\end{multline*}
The expression for the numerator in (\ref{cl1})
can be obtained by simple recursion.
Denote by $l_{k,i}$ the probability
that $k$ consecutive arrivals are lost
and $(k+1)^{st}$ arrival is accepted.
It holds
\begin{eqnarray*}
l_{1,i}=\sum_{k=1}^{i} \delta (1- d_k) (1-\delta)^{i-k}
+(1-\delta)^{i}(1- d_0), \ 1 \le i \le N+1,
\\
l_{n,1}=\delta d_1 l_{n-1,1},
\
l_{n,i}=\delta d_i l_{n-1,i}+(1-\delta)l_{n,i-1},
2 \le i \le N+1.
\end{eqnarray*}
Putting all together, we get the  expression for 
$\mathbf{P}\{ L=k \}$:
$$
\mathbf{P}\{ L=k \}
=
{
\sum\limits_{n=0}^N P_n (1-d_{n}) l_{k,n+1}
\over 
\sum\limits_{n=0}^N P_n (1-d_{n})\sum\limits_{i=1}^{n+1} \delta d_i (1-\delta)^{n+1-i}
}, \ k \ge 1.
$$
Now the moments of the number of consecutive losses can be calculated according
to the definition. 

\section{Option 2}

\subsection{Stationary distribution}

The distributions $P_n$ and $P_n(x)$,
as defined by (\ref{pn}) and (\ref{pnx}),
can be found following the same steps 
in the previous section. The only difference
will be in the expressions for $P_n^+$
of the embedded Markov chain 
$\{ \nu(t), \ t \ge 0\}$.
It is straightforward to see that 
the entries of the transition probability matrix $P=(p_{ij})$
of the embedded Markov chain $\{ \nu(t), \ t \ge 0\}$ 
under \textit{Option 2}
have the form
$$
p_{0j}=p_{1j}=
\begin{cases}
\beta_0, & j=0,\\
\beta_j Q_j + \sum_{k=j}^N \beta_k q_{k-j} +  B_N q_{N-j}, & 1 \le j \le N-1,\\
(q_0 + q_N) B_{N-1}, & j=N,
\end{cases}
$$
$$
p_{ij}=
\begin{cases}
0, & j=0,\\
\sum_{k=j-1}^{N-1} \beta_k q_{k-j+1} +  B_{N-1} q_{N-j}, & 1 \le j \le i-2,\\
\beta_{j-i+1} Q_j + \sum_{k=j-1}^{N-1} \beta_k q_{k-j+1} +  B_{N-1} q_{N-j}, & i-1 \le j \le N-1,\\
(q_{0} +  q_{N})B_{N-i} , & j=N,
\end{cases}
\ 2 \le i \le N.
$$
The matrix $P=(p_{ij})$, just like in the case of \textit{Option 1}, 
does not have any special structure.
So the probabilities $P_n^+$  are found by solving the system
of linear algebraic equations 
$$
{\vec P}^+={\vec P}^+P, \ \
{\vec P}^+ {\vec 1} =1,
$$
where ${\vec P}^+= (P^+_0,\dots,P^+_N)$. 
Now the distributions $P_n$ and $P_n(x)$,
and moments of the number in the system
can be computed using the relations in the previous section.

\subsection{Loss probability}

Let $\pi$ be the probability that the arriving customer (or tagged customer)
will be lost. The expression for $\pi$ under \textit{Option 2}
is more involved than under \textit{Option 1} given by (\ref{ploss}).
This is due to the fact that the accepted customer
may be lost either after the fist service completion or the second etc.
and the chance to be lost varies, depending on the number of
new customers, that arrived between successive service completions.

Let us introduce two quantities:

$\gamma_{ij}$, $1 \le i \le N$, $j \ge 0$, --- probability that the arriving customer
finds $i$ customers in the system and until the next
service completion exactly $j$ new customers arrive 
at the system;

$r_{ij}$, $0\le j \le N-1$, $0 \le i \le N-j-1$, --- probability that the customer waiting in the queue
\textit{will not} be served, if there are $j$ customers in front of it in the queue (excluding the one in server)
and $i$ behind.

Given that $\gamma_{ij}$ and $r_{ij}$ are known,
the loss probability $\pi$  can be computed as
\begin{eqnarray}
\pi
=&& \!\!\!\!\!\!\!\!\!\!
P_{N+1}
+
\sum_{j=0}^{N-i}
\gamma_{ij}
\left (
\sum_{k=0}^{i-2}
q_k r_{j,i-2-k}
+
\sum_{k=i}^{i+j-1}
q_k
+
Q_{j+i}
r_{j,i-2}
\right )
+
\nonumber
\\
&&+
\sum_{j=N-i+1}^{\infty}
\gamma_{ij}
\left (
\sum_{k=0}^{i-2}
q_k r_{N-i,i-2-k}
+
\sum_{k=i}^{N-1}
q_k
+
Q_{N}
r_{N-i,i-2}
\right ).
\label{ploss2}
\end{eqnarray}

Due to the PASTA property of Poisson arrivals,
the expression for $\gamma_{ij}$ simply follows 
from the law of total probability:
\begin{equation}
\gamma_{ij}
=
\int_0^d p_{i}(x) {(\lambda x)^j \over j!} e^{-\lambda x} dx,
\ 1 \le i \le N, \ j \ge 0.
\end{equation}
Again by applying the law of total probability,
we get the relations for the recursive computation of $r_{ij}$, $0\le j \le N-1$, $0 \le i \le N-j-1$:
\begin{eqnarray}
r_{i0}
=
&& \!\!\!\!\!\!\!\!\!\!\!\!
\sum_{m=0}^{N-i-1}
\beta_m 
\sum_{k=1}^{m+i}
q_k
+
\sum_{m=N-i}^{\infty}
\beta_m 
\sum_{k=1}^{N-1}
q_k,
\\
r_{ij}=
&& \!\!\!\!\!\!\!\!\!\!\!\!
\sum_{m=i}^{N-1-j}
\beta_{m-i}
\left (
\sum_{k=0}^{j-1}
q_k r_{m,j-1-k}
\!+\!
\sum_{k=j+1}^{m+j}
q_k
\!+\!
Q_{j+m+1}
r_{j,j-1}
\right )
+
\nonumber
\\
&& \!\!\!\!\!\!\!\!\!\!\!\! +
\sum_{m=N-j-i}^{\infty}
\beta_m
\left (
\sum_{k=0}^{j-1}
q_k r_{N-j-1,j-1-k}
\!+\!
\sum_{k=j+1}^{N-1}
q_k
\!+\!
Q_{N}
r_{N-j-1,j-1}
\right )\!\!.
\end{eqnarray}
The expressions above can be further simplified by computing 
the integrals explicitly, but we will not do it since 
for small and moderate values of $d$, $N$ and $\lambda$
they can be directly used for numerical implementation.

\section{Conclusion}

Although the renovation mechanism is based on completely different idea 
than the RED-type AQMs, as numerical experiments
show, it allows to achieve 
comparable system performance.
Yet the choice of the values of its 
parameters $q_i$ presents certain difficulties.
We are unaware of any analytic way of choosing
$q_i$ and thus we have to resort to
special search algorithms. 
Metaheuristics (like particle swarm optimization)
are also applicable here.

\section*{Acknowledgements}
This work was supported by the Russian Foundation for Basic Research (grant 15-07-03406).

\section*{References}


\bibliography{mybibfile}

\end{document}